\documentclass[preprint,pra,superscriptaddress,longbibliography]{revtex4-1}

\usepackage[normalem]{ulem}
\usepackage[english]{babel}
\usepackage{graphicx}
\usepackage{amssymb}
\usepackage{amsmath}
\usepackage[usenames,dvipsnames]{color}
\usepackage{comment}

\usepackage[symbol*]{footmisc}

\DefineFNsymbolsTM{myfnsymbols}{
  \textasteriskcentered *
  \textdagger    \dagger
  \textdaggerdbl \ddagger
  \textsection   \mathsection
  \textbardbl    \|%
  \textparagraph \mathparagraph
}%
\setfnsymbol{myfnsymbols}

\begin{document}

\title{Weak ergodicity breaking of receptor motion in living cells stemming from random diffusivity}

\author{Carlo Manzo\footnote{carlo.manzo@icfo.es}}
\thanks{These two authors contributed equally to this work}

\author{Juan A. Torreno-Pina}
\thanks{These two authors contributed equally to this work}

\author{Pietro Massignan}

\author{Gerald J. Lapeyre Jr.}
\affiliation{ICFO-Institut de Ci\`encies Fot\`oniques, Mediterranean Technology Park, 08860 Castelldefels (Barcelona), Spain}

\author{Maciej Lewenstein}

\author{Maria F. Garcia Parajo\footnote{maria.garcia-parajo@icfo.es}$\,$}
\affiliation{ICFO-Institut de Ci\`encies Fot\`oniques, Mediterranean Technology Park, 08860 Castelldefels (Barcelona), Spain}
\affiliation{ICREA-Instituci\'o Catalana de Recerca i Estudis Avan\c{c}ats, 08010 Barcelona, Spain}

\date{\today}

\begin{abstract}
Molecular transport in living systems regulates numerous processes underlying biological function. Although many cellular components exhibit anomalous diffusion, only recently has the subdiffusive motion been associated with nonergodic behavior. These findings have stimulated new questions for their implications in statistical mechanics and cell biology. Is nonergodicity a common strategy shared by living systems? Which physical mechanisms generate it? What are its implications for biological function?  Here, we use single particle tracking to demonstrate that the motion of DC-SIGN, a receptor with unique pathogen recognition capabilities, reveals nonergodic subdiffusion on living cell membranes. In contrast to previous studies, this behavior is incompatible with transient immobilization and therefore it can not be interpreted according to continuous time random walk theory. We show that the receptor undergoes changes of diffusivity, consistent with the current view of the cell membrane as a highly dynamic and diverse environment. Simulations based on a model of ordinary random walk in complex media quantitatively reproduce all our observations, pointing toward diffusion heterogeneity as the cause of DC-SIGN behavior. By studying different receptor mutants, we further correlate receptor motion to its molecular structure, thus establishing a strong link between nonergodicity and biological function. These results underscore the role of disorder in cell membranes and its connection with function regulation. Due to its generality, our approach offers a framework to interpret anomalous transport in other complex media where dynamic heterogeneity might play a major role, such as those found, e.g., in soft condensed matter, geology and ecology. 
  
\end{abstract}


\maketitle

\section{Introduction \label{sec:intro}}

Cell function relies heavily on the occurrence of biochemical interactions between specific molecules. Encounters between interacting species are mediated by molecular transport within the cellular environment. A fundamental mode of transport for molecules in living cells is represented by diffusion, a motion characterized by random displacements.
 The quantitative study of diffusion is thus essential for understanding molecular mechanisms underlying cellular function, including target search~\cite{Condamin2007}, kinetics of transport-limited reactions~\cite{Lomholt2007,Condamin2008}, trafficking and signaling~\cite{Choquet2003}.  These processes take place in complex environments, crowded and compartmentalized by macromolecules and biopolymers.  A prototypical example is the plasma membrane, where the interplay of lipids and proteins with cytosolic (e.g., the actin cytoskeleton) and extracellular (e.g., glycans) components generates a highly dynamic and heterogeneous organization~\cite{Kusumi2014}.

 The diffusion of a single molecule $j$, whose position $x_j$ is sampled at $N$ discrete times $t_i = i\Delta t$,  is often characterized by the time-averaged mean-square displacement (T-MSD): 
%
\begin{equation}
\text{T-MSD}(t_{lag}=m\Delta t)=\frac{1}{N-m} \sum_{i=1}^{N-m}  \big(    x_j \left(  t_i + m \Delta t \right) -  x_j \left(  t_i \right)  \big)^2, 
\label{T-MSD}
\end{equation}
%
which for a Brownian particle scales linearly in the time-lag $t_{lag}$.  Application of fluorescence-based techniques to living cells has evidenced striking deviations from Brownian behavior in the nucleus~\cite{Bronstein2009}, cytoplasm~\cite{Weiss2004,Golding2006,Jeon2011,Tabei2013} and plasma membrane~\cite{Saxton1997,Weigel2011}. Indeed, numerous cellular components show anomalous subdiffusion~\cite{Hofling2013}, characterized by a power law dependence of the ${\rm MSD}\sim t^\beta$, with $\beta<1$~\cite{Bouchaud1990, Havlin2002, Klafter2011}. Owing to the implications of molecular transport for cellular function and the widespread evidence of subdiffusion in biology, major theoretical efforts have been devoted to understand its physical origin. Subdiffusion is generally understood to be the consequence of molecular crowding~\cite{Saxton1994} and several models have been developed to capture its main features. In general, subdiffusion can be obtained by models of energetic and/or geometric disorder, such as: (i) the continuous-time random walk (CTRW), i.e., a walk with waiting times between steps drawn from a power law distribution~\cite{Klafter1987}; (ii) fractional Brownian motion, i.e., a process with correlated increments~\cite{Mandelbrot1968}; (iii) obstructed diffusion, i.e., a walk on a percolation cluster or a fractal~\cite{Havlin2002}; (iv) diffusion in a spatially and/or temporally heterogeneous medium~\cite{Cherstvy2013,Massignan2014,Chubynsky2014}. Some of these models have been associated with relevant biophysical mechanisms such as trapping~\cite{Saxton1996}, the viscoelastic properties of the environment~\cite{Ernst2012} or the presence of barriers and obstacles to diffusion~\cite{Weigel2012}.
  
Advances in single particle tracking (SPT) techniques have allowed the recording of long single-molecule trajectories and have revealed very complex diffusion patterns in living cell systems~\cite{Kusumi2014,Saxton1997}. Recently, it has been shown that some cellular components show subdiffusion associated with weak ergodicity breaking (wEB)~\cite{Weigel2011,Jeon2011,Tabei2013}, with the most obvious signatures being the non-equivalence of the T-MSD and  the ensemble-averaged MSD (E-MSD). The experimentally determined ensemble-averaged
 MSD over a time interval $m\Delta t$ is defined by:
%
 \begin{equation}
 \text{E-MSD}(t_{lag}=m\Delta t)=\frac{1}{J} \sum_{j=1}^{J}  \big(     x_j \left( t_i + m \Delta t \right) -  x_j \left( t_i \right) \big)^2, 
 \label{E-MSD}
 \end{equation}
%
where $J$ is the number of observed single-particle
trajectories and $t_i$ is the starting time relative
to first point in the trajectory.

Moreover, ergodicity breaking has been further confirmed by the
presence of aging~\cite{Young1997,Barkai2003}, i.e. the dependence of
statistical quantities on the observation time. Based on these
findings, several stochastic models presenting nonstationary (and thus
nonergodic) subdiffusion have been proposed
~\cite{Barkai2012,Cherstvy2013,metzler2014anomalous,cherstvy2014particle,Jeon2013a}.
Among these, CTRW has been used to model nonergodic subdiffusion in
living cells~\cite{Weigel2011,Jeon2011,Tabei2013} and has begun to
provide theoretical insight into the physical origin of wEB in
biological systems~\cite{Barkai2012}, associating the nonergodic
behavior with the occurrence of particle
immobilization with a heavy-tailed  distribution of trapping times.

  At the same time, these intriguing findings have generated new questions: Is nonergodic subdiffusion a strategy shared by other biological systems? Can biophysical mechanisms other than trapping lead to similar behaviors? What is its functional relevance? Elucidating these issues is crucial to unravel the role of nonergodic subdiffusion in cellular function. The main aim of the present work is to explore other forms of transport in biological systems to provide answers to these questions.

 Here we used SPT to study the diffusion of a prototypical  transmembrane protein, the pathogen-recognition receptor DC-SIGN~\cite{Geijtenbeek2000} on living cell membranes. Our experiments and data analysis show that DC-SIGN dynamics display clear signatures of wEB and aging. However, in contrast to recent studies reporting nonergodic behavior of other membrane proteins~\cite{Weigel2011}, we find that DC-SIGN very rarely shows trapping events so that the observed wEB cannot be described by the CTRW model. Instead, our analysis shows that DC-SIGN displays a heterogeneous dynamics presenting frequent changes of diffusivity. Our numerical simulations, based on a novel theoretical model of Brownian diffusion in complex media~\cite{Massignan2014}, quantitatively reproduce DC-SIGN dynamics demonstrating that nonergodic subdiffusion is a consequence of temporal and/or spatial heterogeneity. Furthermore, structurally mutated variants of DC-SIGN, with impaired function, show very different dynamical features. These results allow us to link receptor transport to molecular structure and receptor function, such as the capability to capture and uptake pathogens.
\begin{center}
\begin{figure}[ht]
\includegraphics[width=0.5\textwidth,clip]{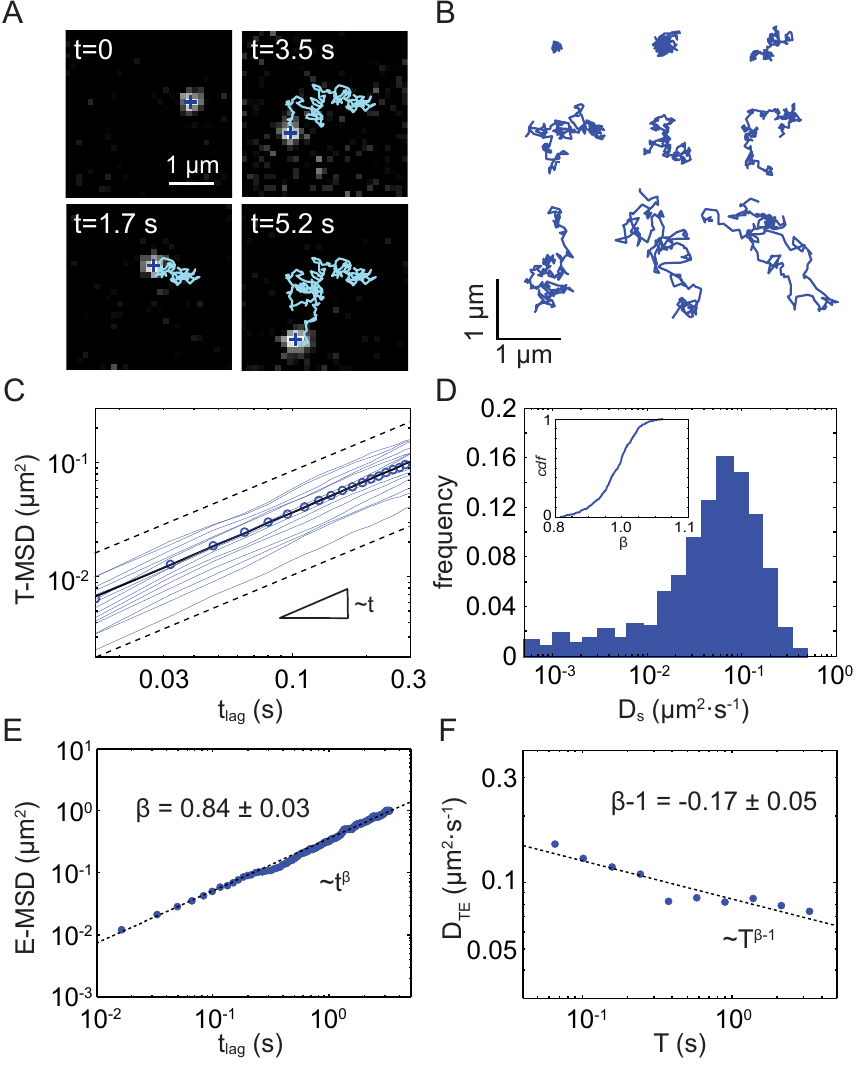}
\caption{{\bf DC-SIGN diffusion shows weak ergodicity breaking and aging.} (A) Representative video frames of a quantum-dot-labeled wtDC-SIGN molecule diffusing on the dorsal membrane of a CHO cell. The centroid position of the bright spot (+), corresponding to a single quantum-dot, is tracked and reconnected to build up the DC-SIGN trajectory, shown by the cyan line. (B) Representative trajectories for the same recording time (3.2 s). (C) Log-log plot of the time-averaged MSD for individual trajectories (blue lines). The dashed lines scale linearly in time, showing that T-MSD is compatible with pure Brownian motion ($\beta=1$).  The symbols ($\bigcirc$) correspond to the average T-MSD. Linear fit to the log-log transformed data (black line) provided $\beta =  0.95 \pm 0.05$. (D) Distribution of short-time diffusion coefficients as obtained from linear fitting of the time-averaged MSD for all the trajectories. (Inset) cdf of the exponent $\beta$ obtained from nonlinear fitting of the T-MSD of all the trajectories. (E) Log-log plot of the ensemble-averaged MSD. Power law fit of the data (dashed line) provides an exponent $\beta=0.84$, showing subdiffusion. (F) Log-log plot of the time-ensemble-averaged diffusion coefficient as a function of the observation time T. The diffusion coefficients are obtained by linear fitting of the time-ensemble-averaged MSD. A power-law fit (dashed line) provides an exponent $\beta-1=-0.17$, revealing aging and in good agreement with the value of $\beta$ found in (E).}
\label{fig:1}
\end{figure}
\end{center}
\section{Weak ergodicity breaking and aging in DC-SIGN dynamics\label{sec:SPT}}
In this work, we performed SPT experiments~\cite{Kusumi2014} to follow the lateral diffusion of the pathogen-recognition receptor DC-SIGN~\cite{Geijtenbeek2000} on living cell membranes. DC-SIGN is a protein exclusively expressed on the surface of cells of the immune system, such as dendritic cells. The receptor is involved in the binding and uptake of a broad range of pathogens such as HIV-1, Ebola virus, hepatitis C virus, \textit{Candida albicans} and \textit{Mycobacterium tuberculosis}~\cite{VanKooyk2003}. Previous studies have shown that DC-SIGN expressed on the membrane of Chinese Hamster Ovary (CHO) cells reproduces the essential features of the receptors naturally occurring on dendritic cells~\cite{Cambi2007,Cambi2009}, thus serving as a valid model system. To characterize its dynamics, we performed video microscopy of quantum-dot labeled DC-SIGN stably transfected in CHO cells in epi-illumination configuration (Fig.\ \ref{fig:1}A-B, see Appendix \ref{app:bio} for details on cell culture and labeling procedures). In order to follow the standard biology nomenclature and to differentiate it from its mutated forms, in this manuscript we refer to the full receptor as the wild-type DC-SIGN (wtDC-SIGN).

 We tracked quantum dot positions with nanometer accuracy by means of an automated algorithm~\cite{Serge2008}. We acquired more than 600 trajectories, all longer than 200 frames with some as long as 2000 frames, at a camera rate of 60 $\rm{frame} \cdot s^{-1}$  to allow the evaluation and the comparison of time and ensemble averaged MSD. The T-MSD of individual trajectories displayed a linear behavior ($\beta \sim 1$), consistent with pure Brownian diffusion (Fig.\ \ref{fig:1}C). The fitting of the average T-MSD provided a value $\beta = 0.95 \pm 0.05$. In addition, the distribution of the exponents $\beta$ obtained by nonlinear fitting of the T-MSDs of the individual trajectories (inset of Fig.\ \ref{fig:1}D) showed an average $\langle \beta \rangle = 0.98\pm 0.06$.   
  
 Since the T-MSD values corresponding to different trajectories were broadly scattered, for each trajectory we calculated the diffusion coefficient $D_s$ by a linear fit of the T-MSD at time lags $<10\%$ of the trajectory duration~\cite{Michalet2010}. As expected, the resulting values of $D_s$ were found to have a very broad distribution, spanning more than two orders of magnitude (Fig.\ \ref{fig:1}D). 

However, in marked contrast with the T-MSD, the E-MSD deviated significantly from linearity, showing subdiffusion with an exponent $\beta=0.84\pm0.03$  (Fig.\ \ref{fig:1}E). The difference between the scalings of T-MSD and E-MSD is a clear signature of wEB~\cite{Bel2005}. To inquire whether DC-SIGN dynamics also exhibits aging, we computed the time-ensemble-averaged MSD (TE-MSD) by truncating the data at different observation times $T$: 
%
\begin{equation}
\text{TE-MSD}(t_{lag}, T)=\frac{1}{J}\frac{1}{\frac{T}{\Delta t}-m} \sum_{i=1}^{\frac{T}{\Delta t}-m}\sum_{j=1}^{J}    \big(    x_j \left(  t_i + m \Delta t \right) -  x_j \left(  t_i \right)  \big)^2, 
\label{TE-MSD}
\end{equation}
%
and extracting the corresponding diffusion coefficient $D_{\rm TE}$ by linear fitting \cite{Michalet2010}.
In systems with uncorrelated increments, it can be shown under rather general assumptions that $D_{\rm TE}\sim T^{\beta-1}$~\cite{Lubelski2008, Massignan2014}. The observed $D_{\rm TE}$ indeed scaled as a power law with an exponent of $-0.17\pm 0.05$ (Fig.\ \ref{fig:1}F), yielding a value of $\beta$ in good agreement with the exponent determined from E-MSD. These results thus demonstrate that wtDC-SIGN dynamics exhibits aging.
\begin{center}
\begin{figure}[ht]
\includegraphics[width=0.5\textwidth,clip]{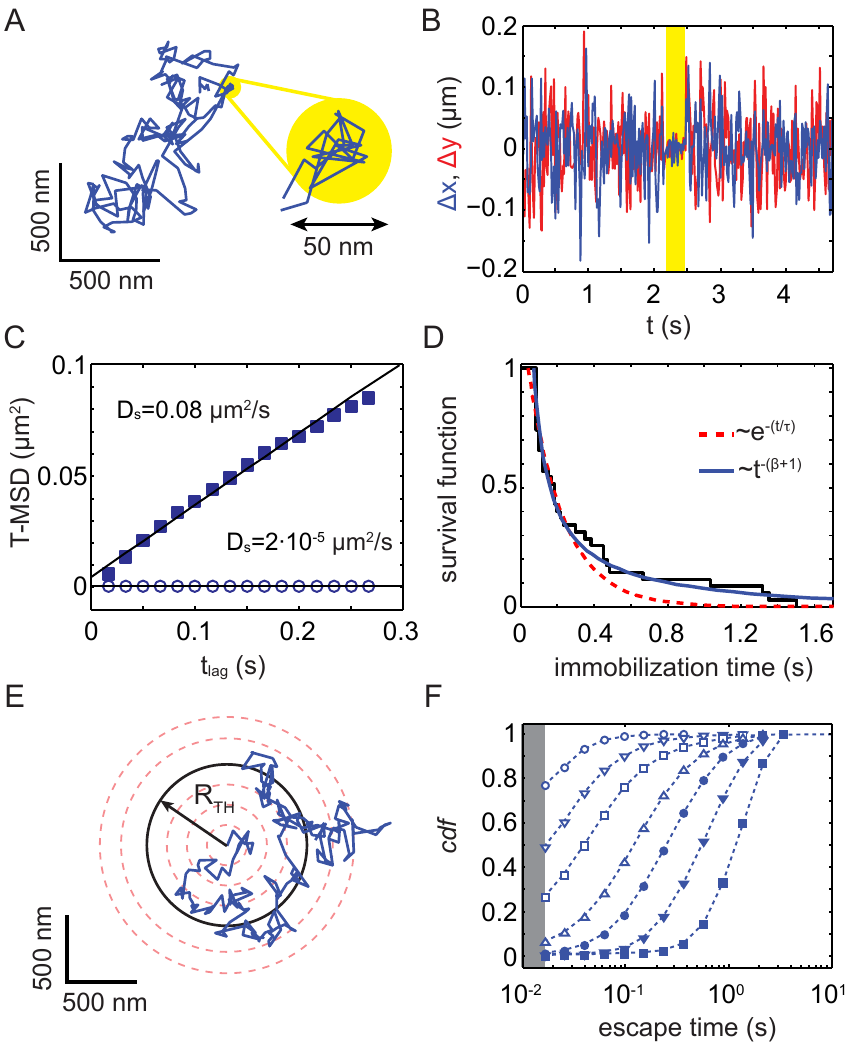}
\caption{{\bf DC-SIGN receptor dynamics is inconsistent with the CTRW model.} (A) A trajectory of wtDC-SIGN on living cell membranes showing a short-lived transient immobilization event, highlighted by the yellow circular area. (B) Plot of the $x$- (blue) and $y$-displacements (red) as a function of time. The occurrence of transient immobilization (yellow region) corresponds to a reduction in the trajectory displacement. (C) Time-averaged MSD for the entire trajectory ($\blacksquare$) and for the immobilization region only ($\bigcirc$). (D) Survival function of the duration of immobilization events for wtDC-SIGN trajectories (black line). Red and blue lines correspond to fits to exponential and power-law distribution functions, respectively. Power law fit provided $\beta = 0.83 \pm 0.05$, in agreement with the exponent obtained for the E-MSD. (E) Schematic representation of the calculation of the escape time probability from circular areas of different radius $R_{\rm TH}$. (F) Cumulative probability distribution function ($cdf$) of trajectory escape time for different radii $R_{\rm TH}$=20 ($\bigcirc$), 50 ($\bigtriangledown$), 100 ($\Box$), 200 ($\bigtriangleup$), 300 ($\bullet$), 500 ($\blacktriangledown$) and to 1000 nm ($\blacksquare$). Dashed lines are guides to the eye. The gray shaded region represents times shorter than the acquisition frame rate.}
\label{fig:2}
\end{figure}
\end{center}
\section{Failure of the CTRW model \label{sec:CTRW}}
 The motion of some biological components, including the Kv2.1 potassium channel in the plasma membrane~\cite{Weigel2011},  lipid granules in yeast cells~\cite{Jeon2011} and insulin-containing vesicles in Pancreatic $\beta$-cells~\cite{Tabei2013}, has been reported to exhibit subdiffusion compatible with the coexistence of an ergodic and a nonergodic process. The nonergodic part of the process has been modeled within the framework of the CTRW~\cite{Bel2005,Barkai2012,Lubelski2008}.
 
CTRW is a random walk in which a particle performs jumps whose lengths have a finite variance, but between jumps the walker remains trapped for random dwell times, distributed with a power-law probability density $\sim t^{-(1+\beta)}$, which for $\beta\leq 1$ has an infinite mean. The duration of trapping events is independent of the previous history of the system. The energy landscape of this process is characterized by potential wells with a broad depth distribution.
 Such energetic disorder yields nonergodicity, since no matter how
 long one measures, deep traps cause dwell times on the order of the
 measurement time. Within the biological context, these traps
 generally have been associated with chemical binding to stationary
 cellular components (e.g. actin cytoskeleton~\cite{Weigel2011} or
 microtubuli~\cite{Tabei2013}), with a distribution of dissociation
 times with varying lifetimes. T-MSDs of molecules performing CTRW
 show broadly scattered values, but are on average linear in the lag
 time $t_{lag}$~\cite{He2008,Lubelski2008}, similar to our observation
 in Fig.\ \ref{fig:1}C. The subdiffusive behavior of the motion
 emerges in the E-MSD, scaling with the same exponent $\beta$
 describing the probability density of trapping dwell-times.

Since DC-SIGN dynamics also showed nonergodic subdiffusion and aging, we sought to investigate whether DC-SIGN diffusion agrees with the predictions of the CTRW model. To this end, we searched for the occurrence of transient trapping events on individual trajectories. 

In SPT experiments, the limited localization accuracy for determining the particle position sets a lower limit for the diffusivity value that can experimentally be measured. In our case, this lower threshold lies at $D_{\rm th}= 6\cdot 10^{-4} \mu {\rm m}^2 {\rm s}^{-1}$. Therefore, a segmentation algorithm~\cite{Montiel2006} was applied to the $x$- and $y$-displacements of our trajectories in order to detect events with diffusivity lower than $D_{\rm th}$. Surprisingly, transient trapping was only detected over less than 5\% of the total recording time (Fig.\ \ref{fig:2}A-C). The detected trapping times displayed an average duration of $330\pm 30$ ms (Fig.\ \ref{fig:2}D). An alternative analysis, based on the transient confinement zone algorithm~\cite{simson1995}, gave comparable results~\cite{Torreno-Pina2014}. 

In order to understand the nature of these trapping events, we attempted to fit their distribution by means of both an exponential and a power law distribution function $\sim t^{-(1+\beta)}$, as expected for CTRW~\cite{Weigel2011}. The power law pdf provided a better fit to the data, yielding an exponent $\beta = 0.83 \pm 0.05$ (Fig.\ \ref{fig:2}D), in agreement with the value obtained for the E-MSD. While a power-law distribution of trapping event durations would be compatible with the behavior expected for the CTRW, it is unlikely that these can have a major role in the ergodicity breaking we observe, given their very small probability of occurrence. In addition, we notice here that various other models predict a similar scaling of the trapping times; an example will be discussed in detail in Sec.\ \ref{sec:suddenChanges}.
 To quantify to what extent the small percentage of trapping events actually influences the nonergodic behavior, we calculated the E-MSD excluding completely the trajectories showing events compatible with immobilization. Interestingly, this analysis provided an exponent $\beta = 0.84 \pm 0.04$ exactly coinciding with the value obtained for the full set of trajectories (Fig.\ \ref{fig:2}D), thus confirming that trapping alone can not account for the ergodicity breaking we observe in wtDC-SIGN dynamics.

 In addition, we constructed the distribution of escape times by identifying the duration of the events in which a trajectory remains within a given radius $R_{\rm TH}$ (Fig.\ \ref{fig:2}E). For a CTRW, the long-time dynamics is dominated by anomalous trapping events and, as a result, this quantity is expected to be independent of $R_{\rm TH}$~\cite{Weigel2011}.  In strong contrast to the CTRW model, the escape-time distributions of DC-SIGN trajectories showed a marked dependence on $R_{\rm TH}$ (Fig.\ \ref{fig:2}F).
 
 In summary, the rare occurrence of transient trapping events, the dependence of escape-time distributions on $R_{\rm TH}$ and, most importantly, the fact that T-MSD and E-MSD show different scaling even when the few trajectories showing immobilization are removed from the analysis, are all inconsistent with CTRW, indicating that the main features of the DC-SIGN dynamics may not be explained in terms of this model.
\begin{center}
\begin{figure}[ht]
\includegraphics[width=0.5\textwidth,clip]{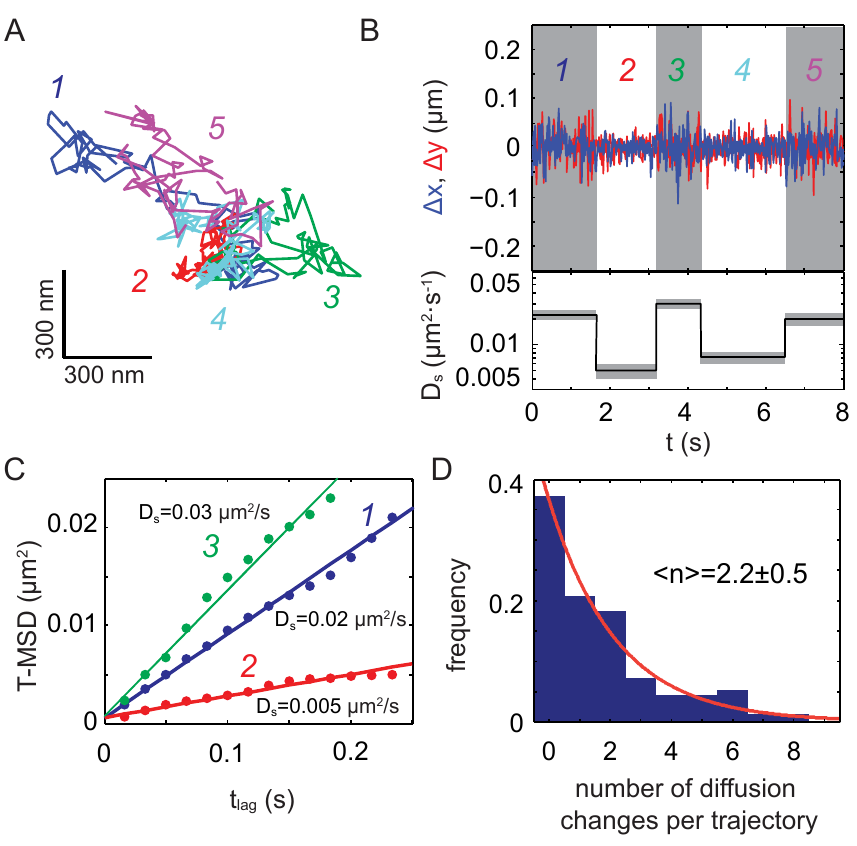}
\caption{{\bf DC-SIGN motion experiences changes in diffusivity.} (A) Representative wtDC-SIGN trajectory displaying changes of diffusivity. Change-point analysis evidenced 5 different regions represented with different colors. (B) Plot of the $x$- (blue) and $y$-displacement (red) for the trajectory in (A) as a function of time. The shaded areas indicate the regions of different diffusivity. The lower panel displays the corresponding short-time diffusion coefficient as obtained from a linear fit of the time-averaged MSD for the 5 different regions. Gray areas correspond to the 95\% confidence level. (C) Plot of time-averaged MSD versus time lag for the first three regions of the trajectory in (A). (D) Histogram of the number of changes of diffusion per trajectory. Most of the trajectories (63\%) display at least one dynamical change, with an average of 2.2 changes per trajectory.}
\label{fig:3}
\end{figure}
\end{center}
\section{DC-SIGN display changes of diffusivity \label{sec:suddenChanges}}
 Recently, diffusion maps of the cell membrane have shown the presence of patches with strongly varying diffusivity~\cite{Serge2008,Cutler2013,Masson2014}. Based on this evidence, we have recently proposed a class of models describing ordinary Brownian motion with a diffusivity that varies randomly, but is constant on time intervals or spatial patches with random size~\cite{Massignan2014}. These models describe anomalous diffusion and wEB in complex and heterogeneous media, such as the cellular environment, without invoking transient trapping. 

 To address whether the observed nonergodic dynamics of DC-SIGN can be described with this theoretical framework, we further analyzed individual trajectories by means of a change-point algorithm to detect variations of diffusivity in time~\cite{Montiel2006}. In brief, the algorithm consists in a likelihood-based approach to quantitatively recover time-dependent changes in diffusivity, based on the calculation of maximum likelihood estimators for the determination of diffusion coefficients and the application of a likelihood ratio test for the localization of the changes. Notably, DC-SIGN trajectories displayed a Brownian motion with relatively constant diffusivity over intervals of varying length, but that changed significantly between these intervals (Fig.\ \ref{fig:3}A-C). Similar features were identified in a large fraction of trajectories, with ~63\% showing at least one diffusivity change (Fig.\ \ref{fig:3}D), in qualitative agreement with the models of random diffusivity~\cite{Massignan2014}.

 To obtain a comprehensive understanding of our data, we considered an annealed model in which randomly diffusing particles undergo sudden changes of diffusion coefficient~\cite{Massignan2014}. The distribution of diffusion coefficients $D$ that a particle can experience is assumed to have a power-law behavior $\sim D^{\sigma -1}$ for small $D$ (with $\sigma>0$) and a fast decay for $D\rightarrow\infty$. Given $D$, the transit time $\tau$ (i.e., the time $\tau$ a particle moves with a given $D$) is taken to have a probability distribution with mean $\sim D^{-\gamma}$ (with $-\infty<\gamma<\infty$). Since the motion during the transit time $\tau$ is Brownian, particles explore areas with radius $r\sim \sqrt{\tau D}$, and the radius of the region explored with such diffusion coefficient has probability distribution with mean $\sim D^{\frac{1-\gamma}{2}}$. Depending on the values of the exponents $\sigma$ and $\gamma$, this model predicts three regimes~\cite{Massignan2014}, namely: (0) for $\gamma<\sigma$, the long-time dynamics is compatible with ordinary Brownian motion and yields an E-MSD exponent $\beta=1$; (I) for $\sigma<\gamma<\sigma+1$, the average transit time $\tau$  diverges and particles undergo nonergodic subdiffusion with $\beta=\sigma/\gamma$; (II) for $\gamma>\sigma+1$, both the average transit time $\tau$ and the average of the radius squared $r^2$ of the explored area diverge and one obtains nonergodic subdiffusion with $\beta=1-1/\gamma$. 
On the other hand, the T-MSD predicted by this model
 remains linear in time for $t\ll T$, for every choice of $\sigma$ and $\gamma$.

\begin{center}
\begin{figure*}[ht]
\includegraphics[width=0.75\textwidth, clip]{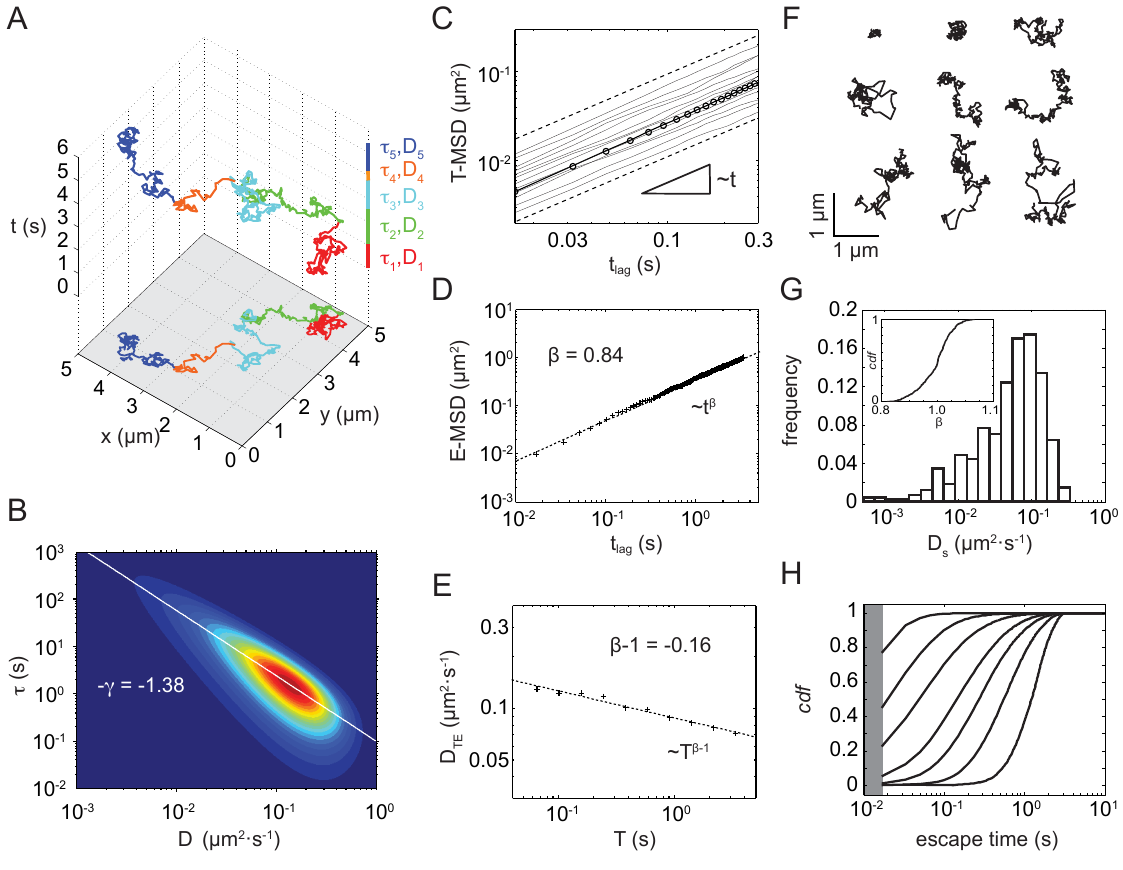}
\caption{{\bf Annealed model of heterogeneous diffusion quantitatively reproduces DC-SIGN motion.} (A) A simulated trajectory composed by five time intervals with different transit time $\tau_i$ and diffusivity $D_i$. (B) Contour plot of the probability distribution of the simulated diffusion coefficient $D$ and transit time $\tau$ for the parameters reproducing the dynamics of wtDC-SIGN ($\sigma=1.16,\,\gamma=1.38,\, b=0.12 \mu{\rm m}^2/{\rm s},\,k=0.10\mu{\rm m}^{2\gamma}{\rm s}^{\gamma+1}$). The white line represents the power law dependence between diffusivity and average transit time with exponent $-\gamma$. (C) Log-log plot of the time-averaged MSD for simulated trajectories (black lines). The symbols ($\bigcirc$) correspond to the average T-MSD. Linear fit to the log-log transformed data provided $\beta =  0.98 \pm 0.03$. (D) Log-log plot of the ensemble-averaged MSD for the simulated trajectories. The dashed line represents a power law with the theoretical exponent $\beta=\sigma/\gamma=0.84$. (E) Log-log plot of the time-ensemble averaged diffusion coefficient as a function of the observation time T. The dashed line represents a power law with the theoretical exponent $\beta-1=-0.16$. (F) Simulated trajectories for the same recording time (3.2 s). (G) Distribution of short-time diffusion coefficients as obtained from linear fitting of the time-averaged MSD for all the simulated trajectories. (Inset) cdf of the exponent $\beta$ obtained from nonlinear fitting of the T-MSD of all the trajectories. (H) $cdf$ of trajectory escape time for different radii. Curves from left to right correspond to radii $R_{\rm TH}$=20, 50, 100, 200, 300, 500 and 1000 nm.}
\label{fig:4}
\end{figure*}
\end{center}

 We performed \textit{in silico} experiments of 2D diffusion (Fig.\ \ref{fig:4}A-B), assuming a distribution of diffusion coefficients $D$ given by:
\begin{equation}
P_D(D)=\frac{D^{\sigma-1}e^{-D/b}}{b^\sigma\Gamma(\sigma)}
\label{diffDistr}
\end{equation}
and a conditional distribution of transit times $\tau$ given by:
\begin{equation}
P_\tau(\tau|D)=\frac{D^\gamma}{k}e^{-\tau D^\gamma/k}
\label{tauDistr}
\end{equation}
where $b$ and $k$ are dimensional constants and $\Gamma(x)$ is the Gamma function. 

The functional forms of the distributions in Eqs.\ \eqref{diffDistr} and \eqref{tauDistr} comply with the requirements of our model, while at the same time ensure the minimal number of free parameters, making them a natural choice for our theoretical
analysis. However, we note here that the asymptotic behavior of the
model is actually independent of the specific functional form of
the joint distributions.
 We performed simulations for different values of $\sigma$, with $\gamma= \sigma/\beta$ as in regime (I), and $\beta=0.84$, the exponent obtained from the experimental E-MSD. The simulations quantitatively reproduce not only subdiffusion, nonergodicity and aging, but also the heterogeneous distribution of diffusion coefficients and escape time distributions (Fig.\ \ref{fig:4}C-H). The remarkable agreement between simulations and experimental data strongly supports heterogeneous diffusion as the origin for DC-SIGN nonergodicity. 

It must be noticed that, in contrast to CTRW, our model does not assume particle immobilization, but a continuous distribution of diffusivity, with $P_D(D) \sim D^{\sigma -1}$ for small $D$. However, from the experimental point of view, it is not possible to distinguish immobilization from very slow diffusion. In fact, the limited localization accuracy of SPT experiments translates into a lower limit for the diffusivity value $D_{\rm th}$ that can be detected. Therefore, in our analysis, trajectories, or portion of trajectories, with diffusivity lower than this threshold value ($D_{\rm th}= 6\cdot 10^{-4} \mu {\rm m}^2 {\rm s}^{-1}$) are identified as immobile, as shown in Fig.\ \ref{fig:2}A-B. From the model described above, the distribution of the duration of these ``apparent'' immobilizations can be calculated as:   
\begin{equation}
P_{\rm imm}(\tau)=\int_{0}^{D_{\rm th}}  P_D(D) P_\tau(\tau|D) dD.
\label{integral}
\end{equation}
We neglect here the possibility that the trajectory of an {\it in-silico} particle contains two consecutive segments characterized by diffusivities $D_{i}$ and $D_{i+1}$ which are both smaller than $D_{\rm th}$, as this probability is vanishingly small for the parameter regime of our setup.
Independently of $D_{\rm th}$, the integral in Eq.\ \eqref{integral} scales asymptotically as $\tau^{-1-\beta}$ with $\beta=\frac{\sigma}{\gamma}$, providing for the distribution of immobilization events the same behavior predicted by the CTRW~\cite{Weigel2011}.  Therefore, the distribution of immobilization times in Fig.\ \ref{fig:2}D is fully compatible with the prediction of our model, further confirming its agreement with the experimental data.

\section{Dynamics of receptor mutants  \label{sec:Mutants}}
From the structural point of view (Fig.\ \ref{fig:5}A),  DC-SIGN is a tetrameric transmembrane protein, with each of the four subunits comprising: (i) an extracellular domain that allows binding of the receptor to pathogens, i.e., ligand binding domain; (ii) a long neck region; and (iii) and a transmembrane part followed by a cytoplasmic tail that allows interactions with inner cell components and facilitates the uptake and internalization of pathogens.~\cite{Feinberg2001, mitchell2001}. Moreover, DC-SIGN contains a single \textit{N}-glycosylation site mediating interactions with glycan-binding proteins~\cite{Torreno-Pina2014}. To gain insight into the molecular mechanisms of DC-SIGN nonergodic diffusion, we generated three mutated forms of the receptor (Fig.\ \ref{fig:5}A). These mutations have been reported to modify the interaction of DC-SIGN with other cellular components, strongly affecting DC-SIGN function~\cite{Smith2007,Manzo2012,Torreno-Pina2014}. The N80A mutant lacks the \textit{N}-glycosylation site. This defect hinders interactions of DC-SIGN with components of the extracellular membrane that bind to sugars~\cite{Torreno-Pina2014}. The $\Delta$35 mutant lacks a significant part of the cytoplasmic tail, preventing interactions with cytosolic components such as actin~\cite{Smith2007}. Finally, the $\Delta$Rep mutant lacks part of the neck region, abrogating interactions between different DC-SIGN molecules~\cite{Manzo2012}.

 We found that each mutation has a very different effect on the dynamics of the receptor. The N80A mutant (Fig.\ \ref{fig:5}C-F) showed nonergodic subdiffusion, with an exponent $\beta$ similar to the one measured for wtDC-SIGN. However, N80A showed a significantly larger extent of heterogeneity in the diffusion coefficients distribution, with a lower median diffusivity. The $\Delta$35 mutant (Fig.\ \ref{fig:5}G-L) also showed nonergodic subdiffusion. The anomalous exponent and the distribution of the diffusion coefficients were similar to that of wtDC-SIGN, with only a slight reduction in median diffusivity. We accurately reproduced N80A and $\Delta$35 dynamics by simulations performed in regime (I), i.e., nonergodic subdiffusion, using comparable values of $\gamma$ for wtDC-SIGN and $\Delta$35, and a smaller value of $\gamma$ for N80A (Fig.\ \ref{fig:5}B). On the other hand, $\Delta$Rep dynamics yielded ergodic Brownian diffusion (Fig.\ \ref{fig:5}M-P) and a narrower distribution of diffusivity with median value significantly higher than for wtDC-SIGN. Consistently, the behavior of $\Delta$Rep was fully captured by \textit{in silico} experiments in regime (0), i.e., ordinary Brownian motion.
 
 Overall, these results demonstrate that the molecular structure of the receptor strongly influences its diffusive behavior on the cell membrane and the occurrence of weak ergodicity breaking. 

\section{Nonergodicity and biological function  \label{sec:Function}}

 Together with our previous biophysical studies on DC-SIGN~\cite{Manzo2012,Torreno-Pina2014}, the data and analysis presented in this paper allow us to link the dynamical behavior of DC-SIGN to its functional role in pathogen capture and uptake (known as endocytosis). 
  
 In terms of steady-state organization, wtDC-SIGN, N80A and $\Delta$35 preferentially form nanoclusters on the cell membrane, which are crucial for regulating pathogen binding~\cite{Manzo2012,Torreno-Pina2014}, whereas removal of the neck region ($\Delta$Rep) reduces nanoclustering and binding efficiency to small pathogens, such as viruses~\cite{Manzo2012}. Our results thus show that the diffusive behavior of the receptor is strongly linked to nanoclustering, but not merely due to size-dependent diffusivity and/or time-dependent cluster formation and breakdown. In fact, dual-color SPT experiments performed at high labeling density do not reveal correlated motion between nearby DC-SIGN nanoclusters, excluding the occurrence of dynamic nanocluster coalescence~\cite{Torreno-Pina2014}. Moreover, although superresolution imaging has revealed that wtDC-SIGN, N80A and $\Delta$35 form nanoclusters with similar distributions of size and stoichiometry~\cite{Manzo2012,Torreno-Pina2014}, our dynamical data evidence significant differences in their diffusion patterns (Figs.\ 1 and 5). 

 Our data are in fact consistent with the view of the plasma membrane as a highly dynamic and heterogeneous medium, where wEB stems from the enhanced ability of DC-SIGN nanoclusters to interact with the membrane environment, including components from the outer and inner membrane leaflet. This interaction is inhibited (or strongly reduced) in the case of the $\Delta$Rep mutant since it does not form nanoclusters~\cite{Manzo2012}. As a result, the motion of $\Delta$Rep is Brownian and ergodic, and interestingly this dynamic behavior correlates with its impaired pathogen binding capability~\cite{Cambi2007,Manzo2012}.
 
\begin{center}
\begin{figure*}[ht]
\includegraphics[width=.75\textwidth, clip]{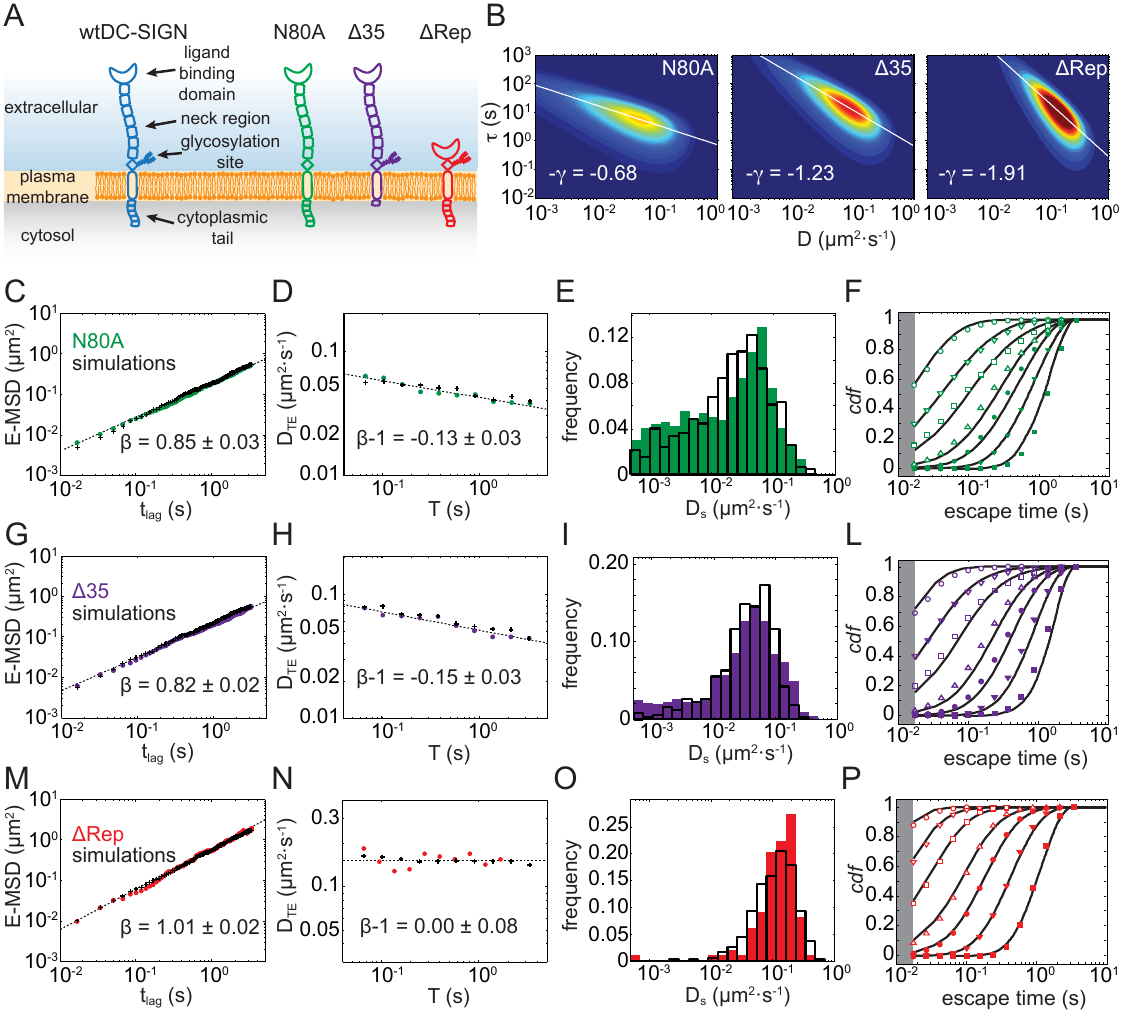}
\caption{{\bf Effect of mutations on the dynamics of DC-SIGN.} (A) Schematic representation of wtDC-SIGN and its mutated forms. (B) Contour plot of the probability distribution of the simulated diffusion coefficient (D) and transit time ($\tau$) for the parameters reproducing the dynamics of N80A ($\sigma=0.58,\,\gamma=0.68,\,b=0.09 \mu{\rm m}^2/{\rm s},\,k=0.74\mu{\rm m}^{2\gamma} {\rm s}^{\gamma+1}$), 
$\Delta$35 ($\sigma=1.04,\,\gamma=1.23,\,b=0.08 \mu{\rm m}^2/{\rm s},\, k=0.07\mu{\rm m}^{2\gamma} {\rm s}^{\gamma+1}$) and 
$\Delta$Rep ($\sigma=2.11,\,\gamma=1.91,\,b=0.07 \mu{\rm m}^2/{\rm s},\, k=0.07\mu{\rm m}^{2\gamma} {\rm s}^{\gamma+1}$). 
The white line represents the power law dependence between diffusivity and average transit time with exponent $-\gamma$. (C) Log-log plot of the ensemble-averaged MSD for N80A trajectories ($\bullet$) and simulated data (+). (D) Log-log plot of the time-ensemble averaged diffusion coefficient for N80A trajectories ($\bullet$) and simulated data (+) as a function of the observation time T. (E) Distribution of short-time diffusion coefficients as obtained from linear fitting of the time-averaged MSD for the N80A (filled bars) and the simulated trajectories (empty bars). (F) $cdf$ of the escape time for N80A (symbols) and simulated trajectories (lines) for different radii. The meaning of the symbols is the same as in Fig.\ \ref{fig:2}F. (G-L)     Dynamical behavior of the $\Delta$35 mutant. (M-P) Dynamical behavior of the $\Delta$Rep mutant.}
\label{fig:5}
\end{figure*}
\end{center}

 In contrast, we observed that both wtDC-SIGN and N80A, which show a similar degree of nanoclustering~\cite{Torreno-Pina2014}, exhibit wEB. But, the distribution of diffusivity of N80A is significantly broader than that of wtDC-SIGN, and is shifted towards lower diffusivity values (Fig.\ \ref{fig:5}C-F). This increased heterogeneity correlates with altered interactions of the N80A with extracellular components, resulting from the removal of the glycosylation site. Indeed, we have recently shown that the N80A mutant has a reduced capability to interact with extra-cellular sugar binding partners~\cite{Torreno-Pina2014}. Thus, it appears that the extracellular milieu next to the membrane contributes to the degree of dynamical heterogeneity sensed by the receptor. Remarkably, this correlation also extends to the functional level, as we have recently shown that interactions of DC-SIGN with extracellular sugar-binding proteins influence encounters of DC-SIGN with the main endocytic protein clathrin. In turn, this resulted in reduced clathrin-dependent endocytosis of the receptor and its pathogenic ligands~\cite{Torreno-Pina2014}. 
 
 Finally, the $\Delta$35 mutant exhibits nanoclustering~\cite{Manzo2012} and wEB similar to that of wtDC-SIGN,
From the biological point of view, however, this mutant is not able to interact with cytosolic components in close proximity to the inner membrane leaflet, including actin~\cite{Smith2007}. Therefore, in contrast to the extracellular influence observed for the N80A mutant, the results obtained for the $\Delta$35 mutant indicate that interactions with the actin cytoskeleton, responsible for the CTRW-like behavior of other proteins~\cite{Weigel2011}, do not play a major role in DC-SIGN wEB. Nevertheless, it should be mentioned that the reduced endocytic capability of the $\Delta$35 could not be uniquely attributed to its dynamic behavior on the cell membrane but rather to its impaired interaction with downstream partners involved in internalization~\cite{Smith2007, Manzo2012}.  

%
%
\section{Conclusions and outlook \label{sec:conc}}
 We have demonstrated that the DC-SIGN receptor displays subdiffusive dynamics, characterized by weak ergodicity breaking and aging. In contrast to other biological systems, receptor trajectories do not show significant evidence of transient immobilization with power-law distributed waiting times. Therefore, its nonergodic behavior cannot be explained in terms of the CTRW model. However, DC-SIGN dynamics is highly heterogeneous, with trajectories often displaying sudden changes of diffusivity.  These features are accurately described by a novel model of ordinary diffusion in complex media, strongly suggesting inhomogeneous diffusivity as the cause of DC-SIGN nonergodic behavior. Comparative analysis of three mutated forms of DC-SIGN evidences the importance of specific regions of the receptor structure, known to mediate interactions with other molecules, in receptor dynamics. Since the mutations of these regions differently impair receptor function, the experiments allowed us to establish the relevance of nonergodicity for the regulation of functional mechanisms, such as capacity for pathogen recognition and internalization. 
 
 The evidence that temporal and/or spatial disorder induces subdiffusion and wEB agrees remarkably well with the current view of the plasma membrane as an extremely complex environment. Here, precise tuning of the spatiotemporal organization of membrane components, in addition to biochemical interactions with molecules in the inner and outer membrane leaflet, orchestrate the triggering of cell signaling pathways.    
A detailed understanding of how these specific interactions occur and affect dynamics is still lacking. Future experiments, involving simultaneous tracking of several proteins by means of multicolor SPT~\cite{Cutler2013} might provide a deeper comprehension of these mechanisms at the molecular level.

 The model used to interpret our data provides a flexible and realistic framework to describe anomalous motion in cell membranes. Although in the present work we have focused our simulations on time-dependent changes of diffusivity, similar conclusions can be obtained assuming a spatial dependence, with constant diffusivity on membrane patches of random size~\cite{Massignan2014}. The current data do not allow discrimination between the two scenarios. The application of techniques that combine dynamic and spatial mapping at high labeling conditions~\cite{Masson2014,Manley2008} would be necessary to verify the occurrence of spatial maps of diffusivity. In addition, numerical simulations of spatial-dependent random diffusivity require the construction of 2-dimensional diffusivity maps consistent with the model's probability densities, which is a non-trivial task. 

 While the work presented here focuses on the cell plasma membrane, we point out that these results have much broader implications. In fact, our model and analysis are very general and can be applied to any diffusive system that shows wEB, in order to investigate the role that heterogeneous diffusivity plays in observed anomalies. Fundamental questions about the nature of anomalous and nonergodic diffusion in disordered media arise in many fields, such as life sciences~\cite{Barkai2012}, soft condensed matter~\cite{Vallee2003,Volpe2014}, ultracold gases~\cite{Lucioni2011,Krinner2013}, geology~\cite{berkowitz2006} and ecology~\cite{Bascompte1998}. Our work provides an alternative conceptual framework and specific tools for answering these questions.
  
\acknowledgments

We thank A. Cambi for providing DC-SIGN transfected CHO cells, O. Esteban for preparing the Fab fragments, P. Symeonidou Besi for preliminary data analysis and B. Castro for recording the N80A trajectories. This work was supported by Fundaci\'o Cellex, Generalitat de Catalunya (Grant No. 2009 SGR 597), the European Commission (FP7-ICT-2011-7, Grant No. 288263), the HFSP (Grant No. RGP0027/2012), ERC AdG Osyris, and the Spanish Ministry of Science and Innovation (Grants FOQUS and No. MAT2011-22887).

\appendix
\section{Cell culture and labeling \label{app:bio}}

Chinese hamster ovary (CHO) cells stably transfected with DC-SIGN mutants were cultured in HAM's F-12 medium (LabClinics) supplemented with 10\% fetal calf serum and antibiotic/antimycotic (Gibco). CHO cells were seeded onto 25 mm coverslips 24 hours before imaging. Streptavidin-coated quantum dots (Qdot655, Invitrogen) were added to an equimolar solution of biotinylated anti-DC-SIGN DCN46 Fab fragment (or anti-AU1 monovalent Ab, in the case of $\Delta$Rep mutant) and a 50x excess of free biotin (Gibco) in order to obtain a 1:1 Fab fragment-quantum dot ratio (or monovalent Ab-quantum dot ratio, in the case of $\Delta$Rep mutant). In these conditions, we estimated a 0.04\% probability of having multiple Fab fragments or Abs bound to the same quantum dot. 
Cells were incubated for 5 min at RT with 50 pM conjugated quantum dots in cold PBS buffer supplemented with 6\% BSA. Extensive washing was performed to remove non-bound conjugated quantum dots before imaging.  In parallel with each experiment, we also performed control experiments by labeling DC-SIGN-negative CHO cells. The lack of quantum dots binding to the DC-SIGN-negative CHO cells confirmed the absence of non-specific binding. Furthermore, the low-concentration labeling conditions were chosen to minimize the probability of having more than one quantum dot within the same region of interest. Fab fragments and monovalent Ab were obtained as described in Refs.~\cite{Manzo2012,Torreno-Pina2014}.

\section{Single particle tracking experiments \label{app:spt}}

We performed video microscopy using a custom single-molecule sensitive epi-fluorescence microscope. Continuous excitation was provided by the 488-nm line of an Argon-ion laser (Spectra Physics), with power density at the sample plane of $\sim$0.3 kW/cm$^2$. Fluorescence was collected by means of a 1.2 NA water immersion objective (Olympus) and guided into an intensified EM-CCD camera (Hamamatsu) after suitable filtering. Movies were recorded on the dorsal membrane of CHO cells at 60Hz frame rate. Experiments were performed in a culture dish incubator (DH-35iL, Warner Instrument) equipped with a temperature controller (TC-324B, Warner Instrument) and a digital CO$_2$ controller (DGTCO2BX, Okolab) at 37$^{\circ}$C and in 5\% CO$_2$ atmosphere. Trajectories were analyzed with custom Matlab code based on the algorithm described in Ref.~\cite{Serge2008}. In order to avoid artifacts in trajectory reconnection caused by quantum dots blinking dynamics and/or high local density of quantum dots, each trajectory was terminated at the first video frame not displaying a clearly identifiable bright spot in the surrounding of the quantum dot localization obtained in the previous frame. Similarly, the trajectory reconstruction was also interrupted if the presence of multiple bright spots did not allow unambiguous identification of the same quantum dots in successive video frames. As a further check for false-positive reconnection, trajectories were overlaid to raw movies and visually inspected.

\section{Data analysis \label{app:data}}

Time-, ensemble- and time-ensemble-averaged mean-squared displacements were calculated as described in~\cite{Weigel2011}. Exponents of the E-MSD, average T-MSD and $D_{\rm t,ens}$ were obtained by linear fitting of the log-log transformed data. Errors were calculated as the 99\% confidence interval of the fitting parameters. Short-time diffusion coefficients were extracted from the linear fit of the first 10\% of the points of T-MSD curves~\cite{Michalet2010}. 

Measurements of the apparent diffusion of quantum dots on fixed cells and glass coverslips were used to estimate the smallest detectable diffusivity. Short-time diffusion coefficients were obtained as described above for trajectories of immobilized quantum dots and the corresponding probability distribution was calculated. 95\% of the immobilized quantum dots trajectories showed values lower than $6\cdot 10^{-4} \mu {\rm m}^2/$s, which was therefore set as the threshold ($D_{\rm th}$) for classifying a trajectory as mobile. 

Dynamical changes in the motion of DC-SIGN receptors were identified by application of the change-point algorithm described in Ref.~\cite{Montiel2006}. In brief, the trajectories were recursively segmented and a maximum-likelihood-ratio test was applied to the trajectory displacements ($\Delta x$, $\Delta y$) in order to identify sudden changes of diffusivity.  The critical values for Type I error rates were set to a confidence level of 99\%, corresponding to 1\% probability of having a false-positive identification of a change-point. For each dynamical region identified by the algorithm, the short-time diffusion coefficient was calculated from a linear fit of the first 10\% of the points of the corresponding MSD curves~\cite{Michalet2010}. Regions showing a short-time diffusion coefficient lower than $D_{\rm th}$ were considered compatible with transient immobilization.

\section{Simulations \label{app:simu}}

Simulated trajectories (500 per parameter set) were obtained by generating random diffusion coefficients $D$ according to the probability distribution given in Eq.\ \eqref{diffDistr}. For each diffusion coefficient, the corresponding transit time $\tau$ was calculated as a random number drawn from the distribution given in Eq.\ \eqref{tauDistr}. Particle coordinates ${\bf r}=\{x,y\}$ were generated as:
\begin{equation*}
\nonumber {\bf r}_{t+\Delta t'}={\bf r}_t+\sqrt{2D\Delta t'}\boldsymbol{\xi}
\end{equation*}
where $\boldsymbol{\xi}=\{\xi_x,\xi_y\}$ are pairs of random numbers from a Gaussian distribution with zero mean and unitary standard deviation. The time increment was calculated as $\Delta t'=\Delta t/n$, where $\Delta t$ is the camera acquisition rate and $n$ is a integer depending on $D$ and $\tau$, chosen in order to have at least 10 points for each interval. For comparison with SPT data, trajectories were sub-sampled at the camera acquisition rate. Simulated trajectories were generated with duration $T_{\rm sim} \geq 3\cdot T_{\rm exp}$, where $T_{\rm exp}$ is the duration of the experimental trajectory. The starting point was randomly drawn from a uniform distribution defined within $0$ and $T_{\rm sim}-T_{\rm exp}$. Trajectories were then cut to have the same duration $T_{\rm exp}$ as the experimental ones. Gaussian noise corresponding to the experimental localization accuracy ($\sigma_{\rm acc}=20$ nm) was subsequently added to the trajectories.

\bibliography{ListShortDot,MS_Manzo_bib}

\end{document}